# Peer review research assessment: a sensitivity analysis of performance rankings to the share of research product evaluated[1]

Giovanni Abramo[a,b,*], Ciriaco Andrea D'Angelo[a], Fulvio Viel[a]

[a] *Laboratory for Studies of Research and Technology Transfer, Department of Management, School of Engineering, University of Rome "Tor Vergata" - Italy*

[b] *National Research Council of Italy*

**Abstract**

In national research assessment exercises that take the peer review approach, research organizations are evaluated on the basis of a subset of their scientific production. The dimension of the subset varies from nation to nation but is typically set as a proportional function of the number of researchers employed at each research organization. However, scientific fertility varies from discipline to discipline, meaning that the representativeness of such a subset also varies according to discipline. The rankings resulting from the assessments could be quite sensitive to the size of the share of articles selected for evaluation. The current work examines this issue, developing empirical evidence of variations in ranking due changes in the dimension of the subset of products evaluated. The field of observation is represented by the scientific production from the hard sciences of the entire Italian university system, from 2001 to 2003.

---

[*] Corresponding author, Dipartimento di Ingegneria dell'Impresa, Università degli Studi di Roma "Tor Vergata", Via del Politecnico 1, 00133 Rome - ITALY, tel/fax +39 06 72597362, abramo@disp.uniroma2.it

# 1. Introduction

A growing number of governments have now identified the evaluation of their national research systems as an essential policy priority. National research evaluations can serve various purposes: encouraging research excellence, optimizing resource allocation; reducing information asymmetry between producers of knowledge and users, and demonstrating that investment in research is effective and delivers public benefits.

Historically, research quality assessments have mostly taken a peer review approach (VTR, 2006; RQF, 2007; RAE, 2008; PBRF, 2008; NRC, 2009), though there has recently been a trend towards adoption of more mixed approaches, known as "informed peer review", with an increased basis in metrics (Geuna and Martin, 2003; Hicks, 2009; REF, 2009; ERA, 2009). However, the role of peer review in national assessments remains dominant, in spite of the fact that there are a number of serious and unresolved issues concerning this approach (Abramo et al., 2009; Moxham and Anderson, 1992; Horrobin, 1990), particularly concerning the subjectivity of a series of judgments involved in the procedure: i) in selecting the peer experts who will evaluate products; ii) in the process applied by the experts, as they evaluate the quality level of products; iii) in the upstream process, applied by the research organizations being examined, as they select the research products that will be submitted for evaluation.

This study addresses a further critical issue that, from the authors' examination of the literature, has not so far been addressed. This issue concerns the fact that peer review exercises are based on observing a subset of the scientific production of the research organizations. Each institution must select a number of products, typically dependant on the number of scientists it employs, for submission to the peer panels. This number varies from nation to nation. For example, in the UK, the most recent Research Assessment Exercise (RAE) considered a subset of four research products for each staff member over a seven year period (representing about 50% of total output). Each panel was to decide which proportion of submitted outputs to examine in detail. For example, in biological sciences panelists choose to examine in detail 25% of submitted outputs; in physics 50%.

For the Research Quality Framework (RQF), in Australia, universities were asked to present work from research groups (minimum 5 members), with each researcher within a group submitting up to 4 research outputs representing a six year period. The guidelines for a new system of assessment in Australia (ERA, 2009) now provide that in the hard sciences (subject to bibliometric analysis) all publications must be included and evaluated. In social sciences, humanities and the arts all publications must be reported by each university, but only a subset (20% of total) is to be evaluated by the expert panels for detailed review. For New Zealand Performance-Based Research Fund (PBRF), each eligible staff member was required to present between one to four research outputs resulting from a six year period, but the panels were then free to choose what sample of these products they would actually evaluate. Overall, about 31% of the submitted research outputs were ultimately examined, but there were large differences among disciplines, from a minimum of examining 15% in the social sciences to a maximum of 100% in the biological sciences. The first Triennial Research Evaluation in Italy (VTR, 2006) limited the subset for consideration to one research product per every four staff members, over a three year period (about 9% of total output). The next Italian exercise, which has just been approved and will cover a five year period, calls for the submission of two products from each staff member. Each expert panel will decide whether to recur to bibliometrics or peer review to evaluate the submitted publications.

Clearly, a limit on the subset of products to be examined permits an evaluation exercise that features reasonable costs and times. The choice to limit the number of products evaluated could also reflect policy objectives of identifying and rewarding excellence, rather than recognizing average quality or productivity of an organization (see the UK's REF). However, one might still be perplexed by the substantial variation in the dimension of the subsets considered by the various national exercises. The question arises: is it best to measure scientific excellence based on 9%, 50%, or on one of the intermediate percentages of research outputs?

In addition, it is well known that "fertility" in scientific publication varies among disciplines, due to variations in the underlying times necessary to mature significant research results[2]. If the rate of selection (number of products to be selected) is based on numbers of research staff, and is homogenous over all the scientific disciplines evaluated, then the subset selected will represent different percentages of the total outputs from the disciplines, with the percentages reaching highs in the less fertile fields and descending to lows for the more fertile fields.

It is clearly difficult to give an answer to the question as to the optimal subset of products for conducting an evaluation. However, it seems that it should be possible, as this work will attempt, to provide an indication of how the performance rankings of research organizations vary with varying dimension of the subset of products considered.

The field for investigation will be all Italian public universities that are active in the hard sciences (a total of 69 institutions), for the 2001-2003 triennium. The following section of the study presents the dataset in detail, and the bibliometric indicators used. Section 3 will examine the representativeness of the subset of products presented to the Italian VTR (2006), by scientific field. Section 4 proposes a bibliometric "simulation" of Italian university rankings, first using the same selection rate of research products as in the VTR (2006), while Section 5 presents a sensitivity analysis of the response of bibliometric rankings to various further scenarios for rate of selection. The closing section of the work offers the authors' comments on the results from the analyses conducted, and their implications.

## 2. Dataset and indicators

The dataset of scientific products examined in the study is derived directly from the Observatory of Public Research (ORP)[3], a database developed and maintained by the authors. The ORP, derived under license from the Thomson Reuters Web of Science (WoS), provides a census of scientific production dating back to 2001, from all Italian public research organizations. For this particular study the analysis is limited to universities. The field of observation covers the 2001-2003 triennium and is limited to the hard sciences, meaning eight out of the total 14 so called "universities disciplinary areas" (UDAs) that comprise the Italian university system. These areas are as follows:

---

[2] For a detailed analysis of the distribution of average productivity among scientific disciplines, see Abramo et al., 2008.
[3] www.orp.researchvalue.it

Mathematics and computer science, Physics, Chemistry, Earth science, Biology, Medicine, Agriculture and veterinary science, Industrial and information engineering[4]. The authors have grouped the WoS hard science subject categories into the 8 UDAs under examination (see Annex in Abramo and D'Angelo, 2009). Thus, the dataset that serves as the basis for the analysis consists of 69,351 publications[5] produced in the 2001-2003 triennium by researchers in the 69 universities of Italy, with each publication[6] classified in one of the WoS categories listed in the above said Annex. The quality of each publication is measured by the citations received. The indicator used is defined as the *Article Impact Index* (AII) and is equal to the ratio of the number of citations (observed as of 31/03/2008) for the article divided by the average number of citations for all the articles of the same year that fall in the same WoS subject categories. A value of 1.40, for example, indicates that the article was cited 40% more often than the average.

## 3. Representativeness of the subset of research products submitted to the Italian VTR

In the most recent Italian evaluation exercise (VTR, 2006) each university was required to submit, for each UDA, a share of research outputs equal to 25% of the number of their full-time equivalent (FTE) researchers on staff in that UDA. The products selected by the universities were forwarded to panels of experts, which then assigned them to at least two reviewers specializing in the scientific field of concern, who were responsible for formulating the qualitative judgments of the products. The opinions were expressed as one of four possible grades: excellent, good, acceptable or lacking. The panels then reexamined the opinions submitted and aggregated the results by university, for each disciplinary area. The completed procedure thus provides ratings that give a research quality overview of all the universities, in each UDA. A total of 13,374 products were evaluated, of which 7,907 were from the eight UDAs considered in the current study. Of these products, 7,513 were indexed in the WoS. Table 1 presents the distribution by UDA. It can be seen that, overall, the articles considered in the VTR compose 8.9% of the total publications produced by Italian universities, as listed in the WoS, but the table also illustrates that the data vary from area to area in a substantial manner. For Physics, the products evaluated represent 4.6% of the total listed, which is a very limited share, particularly if compared to the 15.3% share of Biology or the 21.5% for Agriculture and veterinary science. Such variability is clearly due to the varying intensity of publication that characterizes the different UDAs.

---

[4] Note that these disciplines include 60% of Italy's total university research personnel. Civil engineering and architecture are not considered because the WoS listings are not sufficiently representative of research output in this area.
[5] The publications considered are those referred to in the WoS as "article" or "review", and exclude all other types of publications.
[6] Publications in multidisciplinary journals were distributed to the relevant UDAs.

| University Disciplinary Area | VTR products | Of which WoS publications (a) | Total WoS-listed publications (b) | a/b |
|---|---|---|---|---|
| Mathematics and computer science | 751 | 711 (94.7%) | 6,722 | 10.6% |
| Physics | 626 | 596 (95.2%) | 12,919 | 4.6% |
| Chemistry | 758 | 712 (93.9%) | 8,991 | 7.9% |
| Earth science | 323 | 303 (93.8%) | 3,827 | 7.9% |
| Biology | 1,279 | 1,239 (96.9%) | 8,103 | 15.3% |
| Medicine | 2,644 | 2,574 (97.4%) | 27,577 | 9.3% |
| Agriculture and veterinary science | 617 | 571 (92.5%) | 2,650 | 21.5% |
| Industrial and information engineering | 909 | 807 (88.8%) | 13,500 | 6.0% |
| *Total* | *7,907* | *7,513 (95.0%)* | *84,289* | *8.9%* |

*Table 1: Number of publications selected for the VTR by Italian universities, in each UDA, and their representativeness (period 2001-2003)*

When the situation is examined at the level of individual universities, the share of products is again found to be a highly variable subset (Table 2). For example in Mathematics and computer science, from calculations of the ratio of the publications submitted to the VTR to the total of publications, the proportion varies from a maximum of 50% for the University of Naples "Parthenope" to a minimum of 1.6% for the International School for Advanced Studies of Trieste. These two universities also hold the two extremes in the rankings of share size for Physics publications. In Chemistry, the maximum value for the ratio of publications presented to total publications is seen for the University of Bari (13%), and the minimum is for the University of Verona (1.5%). In Earth science, the University of Cagliari (16.9%) and the Polytechnic University of Milan (1.1%) hold the first and last positions. Similarly substantial ranges of variation are also seen in other UDAs. It is clear that this variability reflects the varying productivity of research staff at each university.

| UDA | University | Research staff (a) | Publications to select (b = a/4) | Total publications (c) | Sampling rate (b/c)*100 |
|---|---|---|---|---|---|
| Mathematics and compter science | University of Naples "Parthenope" | 5 | 1 | 2 | 50.0 |
| | University of Turin | 143 | 36 | 197 | 18.3 |
| | University of Bari | 90 | 23 | 127 | 18.1 |
| | … | | | | |
| | Scuola Normale Superiore in Pisa | 14 | 4 | 111 | 3.6 |
| | University of Bergamo | 5 | 1 | 30 | 3.3 |
| | International School for Advanced Studies of Trieste | 13 | 3 | 182 | 1.6 |
| Physics | University of Naples "Parthenope" | 7 | 2 | 7 | 28.6 |
| | University of Chieti "Gabriele D'Annunzio" | 6 | 2 | 10 | 20.0 |
| | University of Messina | 51 | 13 | 188 | 6.9 |
| | … | | | | |
| | University of Venice "Ca' Foscari" | 5 | 1 | 55 | 1.8 |
| | Scuola Normale Superiore in Pisa | 19 | 5 | 400 | 1.3 |
| | International School for Advanced Studies of Trieste | 28 | 7 | 635 | 1.1 |
| Chemistry | University of Bari | 103 | 26 | 200 | 13.0 |
| | University of Reggio Calabria "Meditteranean" | 4 | 1 | 8 | 12.5 |
| | Polytechnic University of Bari | 7 | 2 | 17 | 11.8 |
| | … | | | | |
| | University of Catanzaro "Magna Grecia" | 4 | 1 | 32 | 3.1 |
| | University of Trento | 8 | 2 | 79 | 2.5 |
| | University of Verona | 4 | 1 | 68 | 1.5 |
| Earth science | University of Cagliari | 52 | 13 | 77 | 16.9 |
| | University of Benevento "Sannio" | 12 | 3 | 18 | 16.7 |
| | University of Chieti "Gabriele D'Annunzio" | 20 | 5 | 35 | 14.3 |
| | … | | | | |
| | University of Venice "Ca' Foscari" | 8 | 2 | 90 | 2.2 |
| | Polytechnic University of Ancona | 6 | 1 | 71 | 1.4 |
| | Polytechnic University of Milan | 4 | 1 | 88 | 1.1 |
| Biology | University of Naples "Parthenope" | 4 | 1 | 2 | 50.0 |
| | University of Cagliari | 114 | 28 | 114 | 24.6 |
| | University of Messina | 132 | 33 | 142 | 23.2 |
| | … | | | | |
| | University of Teramo | 5 | 1 | 26 | 3.8 |
| | University of Basilicata | 6 | 2 | 56 | 3.6 |
| | International School for Advanced Studies of Trieste | 5 | 1 | 75 | 1.3 |
| Medicine | University of Messina | 492 | 123 | 632 | 19.5 |
| | University of Palermo | 364 | 91 | 604 | 15.1 |
| | University of Catania | 372 | 93 | 710 | 13.1 |
| | … | | | | |
| | University of Camerino | 8 | 2 | 125 | 1.6 |
| | University of Milan "Vita-Salute San Raffaele" | 33 | 8 | 587 | 1.4 |
| | University of Salerno | 4 | 1 | 80 | 1.3 |
| Agriculture and veterinary science | University of Venice "Ca' Foscari" | 4 | 1 | 1 | 100.0 |
| | University of Palermo | 120 | 30 | 60 | 50.0 |
| | University of Sassari | 126 | 32 | 75 | 42.7 |
| | … | | | | |
| | University of Modena and Reggio Emilia | 5 | 1 | 31 | 3.2 |
| | University of Siena | 5 | 1 | 36 | 2.8 |
| | University of Cagliari | 6 | 1 | 49 | 2.0 |
| Industrial and information engineering | University of Castellanza "Carlo Cattaneo" | 6 | 2 | 2 | 100.0 |
| | Academic institute of Architecture in Venezia | 7 | 2 | 5 | 40.0 |
| | Polytechnic University of Bari | 121 | 30 | 170 | 17.6 |
| | … | | | | |
| | University of Milan | 18 | 5 | 498 | 1.0 |
| | University of Verona | 5 | 1 | 114 | 0.9 |
| | University of Milan "Bicocca" | 5 | 1 | 271 | 0.4 |

*Table 2: Representativeness of publications selected for the VTR (2001-2003), in each UDA, by individual universities (top three and lowestl three in the rank are reported).*

## 4. Bibliometric simulation of Italian university rankings based on the 2006 VTR product share

In the preceding section we saw that the dimension of the subset submitted for evaluation in the Italian VTR was based on a selection rate equivalent to one product for every 4 FTE researchers, which resulted in variable percentages of the total publications, by area and from university to university. Now we ask: what would have happened if the selection rate required of the universities had been different? Would the final rankings have remained the same? The only means to address this question is first to attempt to reproduce the evaluation exercise for 2001-2003, but on a bibliometric basis, thus permitting further reproduction of the exercise, but with varying selection rates. In a preceding work, Abramo et al. (2009) demonstrated the existence of a significant and strong correlation between ratings produced by the VTR peer review (2006) and those obtained from bibliometric indicators of quality. In 1997, Oppenheim had likewise demonstrated the existence of a correlation between citation counts and the 1992 RAE ratings for British research in genetics, anatomy and archaeology. In successive works, similar correlations were also confirmed for the areas of archeology, in the UK (Oppenheim and Norris, 2003), and partially observed by Rinia et al. (1998) for research programs in condensed matter physics, in the Netherlands.

In this section, the bibliometric exercise will be conducted according to the selection rate originally applied in the Italian VTR. Thus, for each university, we will select a number of publications equal to 25% of the number of FTE researchers. To identify the best products, we will use the bibliometric indicator AII presented in section 2. Once the best publications of each university are identified, we will group all those for each UDA and assign each publication a rating: 1, if it falls in the first quartile of the national distribution for the indicator; 0.8 if it falls in the second quartile; 0.6 if it falls in the third quartile; 0.2 if it falls in the last quartile. By this means we replicate the mechanism of experts assigning grades to the products presented for the VTR. At this point it is then possible to calculate the average rating per university and it relative national ranking[7]. As an example, Table 3 presents the rankings for the universities in the Physics UDA.

---

[7] The analysis here and in the next section of the study excludes, for each UDA, those universities with less than an average of 5 researchers on staff over the triennium considered.

| University | Research staff | Rank |
|---|---|---|
| Scuola Normale Superiore in Pisa | 19 | 1 |
| International School for Advanced Studies of Trieste | 28 | 1 |
| University of Brescia | 11 | 3 |
| University of Varese "Insubria" | 23 | 3 |
| University of "Roma Tre" | 46 | 5 |
| University of Trento | 42 | 6 |
| University of Rome "La Sapienza" | 185 | 7 |
| Polytechnic University of Bari | 12 | 8 |
| University of Salerno | 38 | 9 |
| University of Ferrara | 50 | 10 |
| University of Trieste | 64 | 11 |
| Polytechnic University of Turin | 39 | 12 |
| University of Venice "Ca' Foscari" | 5 | 13 |
| University of Urbino "Carlo Bo" | 9 | 13 |
| University of L'Aquila | 46 | 13 |
| University of Cagliari | 49 | 16 |
| Polytechnic University of Milan | 44 | 17 |
| University of Florence | 111 | 18 |
| University of Perugia | 41 | 19 |
| University of Lecce "Salento" | 64 | 20 |
| University of Catania | 87 | 21 |
| University of Pisa | 92 | 22 |
| University of Padua | 144 | 23 |
| University of Bari | 67 | 24 |
| University of Pavia | 77 | 25 |
| University of Milan "Bicocca" | 59 | 26 |
| University of Sassari | 5 | 27 |
| Sacred Heart Catholic University | 13 | 27 |
| University of Modena and Reggio Emilia | 39 | 27 |
| University of Genova | 93 | 27 |
| University of Milan | 101 | 31 |
| University of Bologna | 123 | 32 |
| University of Turin | 90 | 33 |
| University of Rome "Tor Vergata" | 84 | 34 |
| University of Eastern Piedmont "A. Avogadro" | 15 | 35 |
| University of Camerino | 23 | 36 |
| University of Naples "Federico II" | 146 | 37 |
| Second University of Naples | 9 | 38 |
| University of Basilicata | 13 | 38 |
| Polytechnic University of Ancona | 12 | 40 |
| University of Udine | 13 | 40 |
| University of Parma | 72 | 42 |
| University of Calabria | 36 | 43 |
| University of Messina | 51 | 44 |
| University of Palermo | 62 | 45 |
| University of Benevento "Sannio" | 4 | 46 |
| University of Verona | 5 | 46 |
| University of Chieti "Gabriele D'Annunzio" | 6 | 46 |
| University of Naples "Parthenope" | 6 | 46 |
| University of Siena | 15 | 46 |

*Table 3: Ranking of Italian universities for the Physics UDA by average impact of the best publications produced, with 1 publication selected for every 4 researchers*

## 5. Sensitivity analysis of performance rankings to product share

In this section we will repeat the bibliometric exercise of Section 4 but with different selection rates, to examine the extent of the shifts in ranking under various scenarios. We will begin by considering a selection rate that varies among the areas, but which achieves a representivity for each area that is equal to the overall VTR rate of 8.9%. This procedure maintains the number of total products to be evaluated at 7,513, exactly as in the true VTR, simulating the same overall budget for the exercise. In this case, however, the publications are now divided among the various areas on the basis of their fertility. Table 4 presents the data under this new scenario of subsets: beginning from the total publications listed in the WoS (second column), the calculations then give the numbers to be selected for the exercise (third column); once the numbers of research staff are noted (fourth column) it is now possible to arrive at the theoretical selection rate (products per researcher, fifth column) necessary to obtain a consistent level of total product share.

We next repeat the evaluation exercise illustrated in Section 2, in each area. We extract the best publications of each university, applying the theoretical selection rate indicated in Table 4. We then repeat the remaining steps of the exercise and arrive at new rankings. As an example, Table 5 presents the comparison to the preceding results, for the UDAs of Physics and Biology. Strong correlations are seen (correlation indexes for rankings under the two scenarios of 0.911 for physics and 0.928 for Biology) but there are also variations that can not be ignored. There are a high number of universities that change positions under the two rankings: 40 out of 50 for Physics and 45 out of 53 for Biology. The mean and median shifts in rankings are respectively 4.3 and 3.0, for both areas. The maximum shifts in position are 15 for Physics and 22 for Biology.

| University Disciplinary Area | Total publications (a) | Publications to select (c=8.9%*a) | Research staff (b) | Sampling rate (c : b) |
|---|---|---|---|---|
| Mathematics and computer science | 6,722 | 599 | 3,069 | 1 : 5.1 |
| Physics | 12,919 | 1,152 | 2,508 | 1 : 2.2 |
| Chemistry | 8,991 | 801 | 3,139 | 1 : 3.9 |
| Earth science | 3,827 | 341 | 1,281 | 1 : 3.8 |
| Biology | 8,103 | 722 | 4,827 | 1 : 6.7 |
| Medicine | 27,577 | 2,458 | 10,452 | 1 : 4.3 |
| Agriculture and veterinary science | 2,650 | 236 | 2,946 | 1 : 12.5 |
| Industrial and information engineering | 13,500 | 1,203 | 4,335 | 1 : 3.6 |
| Total | 84,289 | 7,513 | 32,556 | 1 : 4.3 |

Table 4: *Theoretical selection rate descending from a uniform percentage of total output in each UDA*

| Statistics | Physics | Biology |
|---|---|---|
| Correlation | 0.911 | 0.928 |
| Number of variations | 40 (out of 50) | 45 (out of 53) |
| Mean variation | 4.3 | 4.3 |
| Median variation | 3 | 3 |
| Maximum variation | 15 | 22 |

Table 5: *Statistics for variations in ranking of Italian universities in the Biology and Physics UDAs under two different hypotheses for selection of best products*

Thus, although we see strong correlation in the rankings determined under the two selection rates, the analysis also shows many and sizeable variations: the rankings are not insensitive to the size of the subset submitted for evaluation. The choice of selection rate has consequences relative to the objectives of the evaluation exercise and the subsequent possible allocations of funding. We will attempt to further demonstrate this assertion through a sensitivity analysis: we will simulate eight scenarios, each one using a different dimension for the subset of best products to be evaluated. Table 6 again presents the example of the Physics area.

| Scenario | Subset size (% of total publications) | Publications per researcher |
|---|---|---|
| 1 | 4.6 | 1 : 4 |
| 2 | 8.9 | 1 : 2.2 |
| 3 | 10 | 1 : 1.9 |
| 4 | 20 | 1 : 1.1 |
| 5 | 30 | 1 : 0.6 |
| 6 | 40 | 1 : 0.5 |
| 7 | 50 | 1 : 0.4 |
| 8 | 60 | 1 : 0.3 |

*Table 6: Scenarios for sensitivity analysis, Physics UDA*

For each scenario, the bibliometric exercise is repeated as previously illustrated, leading up to a set of rankings. Table 7 presents, for the Physics area, a frequency matrix for the rankings obtained under the eight simulated scenarios. The analysis shows a consistent "volatility" in rankings: the tails remain quite stable (the Advanced Schools of Pisa and Trieste and university of Rome "Tre" always hold the top rankings and the universities of Benevento and Naples "Parthenope" are always in the last decile) but it is also true that many universities demonstrate very substantial variations in rank under the varying scenarios. The University of Verona, for example, places in the first decile in 50% of cases (four out of the eight scenarios considered), but ranks in the second decile under two other scenarios, while in two other cases it ranks as low as the fifth and seventh deciles. The University of Brescia obtains rankings that oscillate between $3^{rd}$ and $38^{th}$ position for the nation, placing once in the first decile, twice in the second, once in the fifth and four times in the eighth. The dispersion in rankings reflects that in the quality of research outputs. The higher the variation in quality, the higher the dispersion in rankings. Higher dispersion seems to occur more frequently with younger and smaller universities, where the research staff is evidently less homogenous in terms of performance.

| University | \multicolumn{10}{c}{Ranking decile} |
| | 1 | 2 | 3 | 4 | 5 | 6 | 7 | 8 | 9 | 10 |
|---|---|---|---|---|---|---|---|---|---|---|
| Scuola Normale Superiore in Pisa | 8 | | | | | | | | | |
| International School for Advanced Studies of Trieste | 8 | | | | | | | | | |
| University of Venice "Ca' Foscari" | 6 | 1 | 1 | | | | | | | |
| University of "Roma Tre" | 8 | | | | | | | | | |
| University of Verona | 4 | 1 | | | 1 | | 1 | | | 1 |
| University of Varese "Insubria" | 3 | 5 | | | | | | | | |
| University of Trento | 2 | 6 | | | | | | | | |
| Polytechnic University of Turin | | 7 | 1 | | | | | | | |
| University of Rome "La Sapienza" | | 7 | 1 | | | | | | | |
| Polytechnic University of Milan | | 2 | 5 | 1 | | | | | | |
| Second University of Naples | | 2 | 2 | 1 | | 1 | 1 | 1 | | |
| University of Camerino | | 3 | 3 | 1 | | | 1 | | | |
| Polytechnic University of Bari | | 3 | 2 | 2 | 1 | | | | | |
| University of Pisa | | | 7 | 1 | | | | | | |
| University of Milan "Bicocca" | | | 5 | 1 | 1 | 1 | | | | |
| University of Ferrara | | 1 | 3 | 2 | 1 | 1 | | | | |
| University of Florence | | | | 8 | | | | | | |
| University of Trieste | | 2 | 5 | 1 | | | | | | |
| University of Rome "Tor Vergata" | | 1 | 3 | 2 | 1 | 1 | | | | |
| University of Salerno | | 1 | 2 | 1 | 4 | | | | | |
| University of Padua | | | | 3 | 5 | | | | | |
| University of Perugia | | | | 4 | 4 | | | | | |
| Polytechnic University of Ancona | | 2 | 1 | 2 | 1 | | | 2 | | |
| University of Modena and Reggio Emilia | | | 1 | 4 | 3 | | | | | |
| University of Milan | | | | 4 | | 3 | 1 | | | |
| University of Bologna | | | | 1 | 3 | 4 | | | | |
| University of Turin | | | | | 4 | 2 | 2 | | | |
| University of Pavia | | | | 1 | 7 | | | | | |
| University of Udine | | | | 1 | 3 | 2 | 2 | | | |
| University of Catania | | | | 3 | 2 | 3 | | | | |
| University of L'Aquila | | | 1 | | 3 | 4 | | | | |
| University of Basilicata | | | | | 3 | 1 | 2 | 2 | | |
| University of Genoa | | | | | 1 | 5 | 2 | | | |
| University of Eastern Piedmont "A. Avogadro" | | | | 1 | | 4 | 3 | | | |
| University of Lecce "Salento" | | | | 1 | 2 | 4 | 1 | | | |
| University of Brescia | | 1 | 2 | | 1 | | 4 | | | |
| Sacred Heart Catholic University | | | 2 | | 1 | 3 | 2 | | | |
| University of Urbino "Carlo Bo" | | 1 | 2 | | 1 | | 4 | | | |
| University of Naples "Federico II" | | | | | | | | 8 | | |
| University of Cagliari | | | | 2 | | 1 | 4 | 1 | | |
| University of Bari | | | | | 1 | | 1 | 1 | 5 | |
| University of Parma | | | | | | | | 8 | | |
| University of Calabria | | | | | | | | 5 | 3 | |
| University of Messina | | | | | | | | 5 | 3 | |
| University of Palermo | | | | | | | | 7 | 1 | |
| University of Chieti "Gabriele D'Annunzio" | | | | | | | 1 | 2 | 5 | |
| University of Siena | | | | | | | | 1 | 7 | |
| University of Sassari | | | | | 1 | | | 3 | 4 | |
| University of Benevento "Sannio" | | | | | | | | | | 8 |
| University of Naples "Parthenope" | | | | | | | | | | 8 |

*Table 7: Frequency matrix of university ranking, in deciles, under eight scenarios of publication share in Physics*

The same analysis was repeated for the Biology UDA: Table 8: Scenarios for sensitivity analysis, Biology UDATable 8 presents the eight scenarios, while Table 9 presents the frequency matrix for the rankings obtained under these scenarios.

| Scenario | Subset size (% of total publications) | Publications per researcher |
|---|---|---|
| 1 | 8.9 | 1 : 6.7 |
| 2 | 10 | 1 : 6 |
| 3 | 15 | 1 : 4 |
| 4 | 20 | 1 : 3 |
| 5 | 30 | 1 : 2 |
| 6 | 40 | 1 : 1.5 |
| 7 | 50 | 1 : 1.2 |
| 8 | 60 | 1 : 1 |

*Table 8: Scenarios for sensitivity analysis, Biology UDA*

The Biology matrix shows results which are quite similar to the Physics area. Despite the stability in the tails (the same four universities hold the top rankings in each scenario, while the same three rank always in the last decile), middlle rankings show a consistent volatility. For example, University of "Roma Tre" places in six different deciles under the eight simulated scenarios. University of Basilicata places in the first decile in 50% of cases, but in other two scenarios its quality rating places it in the sixth and seventh decile.

What then would be the most appropriate selection rate for peer review exercises? There is probably no definitive answer to this question, in all its aspects, but the economic optimum would obviously be a "just" compromise between budgetary needs on the one hand and robustness of final rankings on the other. The solution calls for the optimization of a trade-off involving two related considerations: costs are expected to become more onerous, in substantially linear terms, with increasing dimension of the subset of products to be evaluated (beyond a certain dimension, the fixed costs of the assessment will have only a marginal effect on the total); meanwhile the reliability of the rankings can expected to increase favorably with the dimension of the subset, up to a threshold beyond which it would be held that the further research products submitted are no longer "excellent".

To examine the variability of rankings with variation in the share of product submitted, we now compare the previous eight scenarios against a new reference scenario: an evaluation of all scientific production by the universities (rather than the previous comparison against a subset of production). Figure 1 presents the trends in the coefficient of correlation and the median variation in rankings of each scenario from the benchmark scenario of all production. It is readily apparent that there is a convergence of rankings with increasing dimension of the subset, but that beyond the "30% scenario" the increments in subset achieve only marginal increases in correlation to the evaluation of all the publications. The same analysis was repeated for Biology: in this UDA, the stabilization effect of increasing subset dimension is slightly less evident compared to that seen in Physics: the correlation with the reference scenario seems to increase approximately in a linear manner over the whole spectrum of scenarios considered (Figure 2).

|  | Ranking decile | | | | | | | | | |
| University | 1 | 2 | 3 | 4 | 5 | 6 | 7 | 8 | 9 | 10 |
|---|---|---|---|---|---|---|---|---|---|---|
| International School for Adavanced Studies of Trieste | 8 | | | | | | | | | |
| University of Milan "Vita-Salute San Raffaele" | 8 | | | | | | | | | |
| Scuola Normale Superiore in Pisa | 8 | | | | | | | | | |
| University of Verona | 8 | | | | | | | | | |
| University of Padua | 6 | 2 | | | | | | | | |
| University of Viterbo "Tuscia" | 4 | 2 | | 1 | 1 | | | | | |
| University of Basilicata | 4 | 2 | | | | 1 | 1 | | | |
| University of Udine | 1 | 5 | 2 | | | | | | | |
| University of Teramo | 1 | 5 | 1 | 1 | | | | | | |
| University of Eastern Piedmont "A. Avogadro" | | 8 | | | | | | | | |
| University of Bari | | 5 | 3 | | | | | | | |
| University of Milan | | 5 | 2 | 1 | | | | | | |
| University of Rome "Tor Vergata" | | 4 | 2 | 2 | | | | | | |
| University of Genova | | 3 | 4 | 1 | | | | | | |
| University of "Roma Tre" | | 3 | 1 | | 1 | 1 | 1 | | 1 | |
| University of Siena | | 2 | 4 | 2 | | | | | | |
| University of Benevento "Sannio" | | 2 | 4 | 1 | 1 | | | | | |
| University of Lecce "Salento" | | 2 | 2 | 3 | | 1 | | | | |
| University of Trieste | | 2 | | 4 | 2 | | | | | |
| Polytechnic University of Ancona | | | 7 | 1 | | | | | | |
| University of Bologna | | | 6 | 2 | | | | | | |
| University of Turin | | | 3 | 3 | 2 | | | | | |
| University of Florence | | | 2 | 2 | 3 | 1 | | | | |
| Sacred Heart Catholic University | | | 1 | 6 | 1 | | | | | |
| University of Calabria | | | | 5 | 2 | 1 | | | | |
| University of Catanzaro "Magna Grecia" | | | | 3 | 5 | | | | | |
| University of Rome "La Sapienza" | | | | 2 | 4 | 2 | | | | |
| University of Ferrara | | | | 1 | 3 | 4 | | | | |
| University of Brescia | | | | 1 | 1 | 5 | 1 | | | |
| University of Naples "Federico II" | | | | | 3 | 4 | 1 | | | |
| University of L'Aquila | | | | | 3 | 1 | 4 | | | |
| Second University of Naples | | | | | 2 | 2 | 4 | | | |
| University of Chieti "Gabriele D'Annunzio" | | | | | 1 | 5 | 2 | | | |
| University of Modena and Reggio Emilia | | | | | 1 | 4 | 3 | | | |
| University of Pisa | | | | | 1 | 2 | 4 | 1 | | |
| University of Pavia | | | | | 1 | | 4 | 3 | | |
| University of Milan "Bicocca" | | | | | | 4 | 4 | | | |
| University of Salerno | | | | | | 2 | 2 | 4 | | |
| University of Perugia | | | | | | 1 | 3 | 4 | | |
| University of Foggia | | | | | | 1 | 3 | 1 | 2 | 1 |
| University of Varese "Insubria" | | | | | | 1 | | 7 | | |
| University of Molise-Campobasso | | | | | | 1 | 2 | | 3 | 2 |
| University of Sassari | | | | | | | 5 | 3 | | |
| University of Venice "Ca' Foscari" | | | | | | | 5 | 1 | 2 | |
| University of Catania | | | | | | | 3 | 5 | | |
| University of Urbino "Carlo Bo" | | | | | | | 2 | 6 | | |
| University of Messina | | | | | | | 1 | 7 | | |
| University of Cagliari | | | | | | | 1 | 1 | 6 | |
| University of Parma | | | | | | | | 8 | | |
| University of Palermo | | | | | | | | 3 | 5 | |
| University of Naples "Parthenope" | | | | | | | | | | 8 |
| University of Rome "Foro Italico" | | | | | | | | | | 8 |
| University of Camerino | | | | | | | | | | 8 |

*Table 9: Frequency matrix of university ranking, in deciles, under eight scenarios of publication share in Biology*

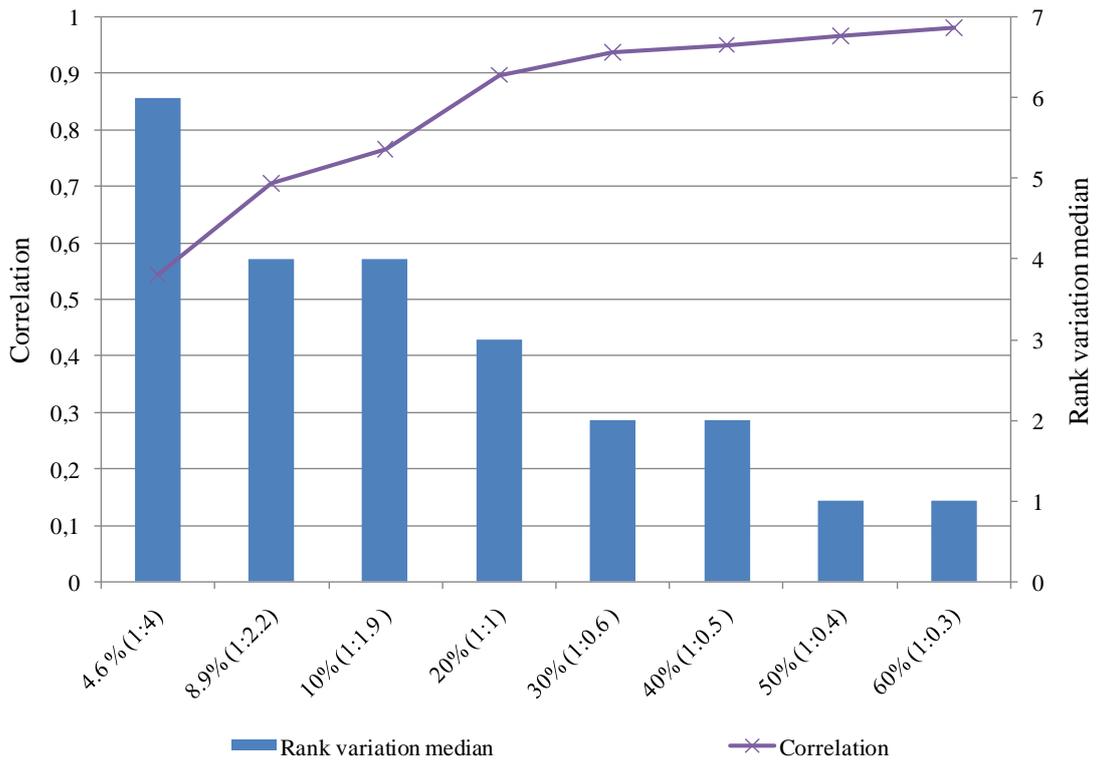

*Figure 1: Trends in the coefficient of correlation and median for variation of rank under various scenarios for product share, for the Physics UDA, with the reference scenario being the evaluation of all products*

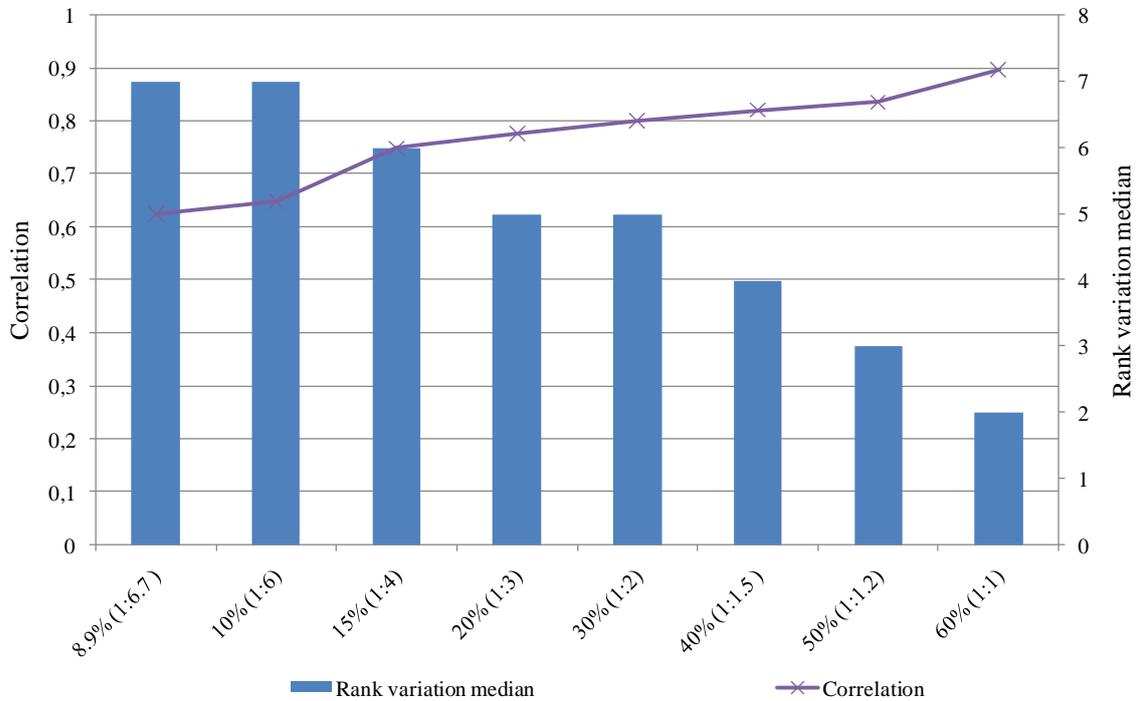

*Figure 2: Trends in the coefficient of correlation and median for variation of rank under various scenarios for product share, for the Biology UDA, with the reference scenario being the evaluation of all products*

# 6. Conclusions

In a number of national research assessment exercises, the products submitted for peer review evaluation represent a percentage of research output that varies with each discipline. The subset of products to be submitted for evaluation is often proportionate to the number of research staff per discipline. But the fertility of the disciplines, meaning the number of research products produced on average per researcher, varies from discipline to discipline, since the time to mature useful scientific results (and the consequent intensity of publications) is greater in some disciplines than in others. The authors posed the question as to whether it could be opportune to consider a selection rate that varies according to discipline, so that the products selected would always represent the same percentage of the total publications of each discipline. A bibliometric simulation of this scenario revealed that it produces substantial shifts in performance rankings with respect to the rankings obtained under a selection rate that remains fixed for all disciplines. A subsequent sensitivity analysis conducted for eight different scenarios of selection rate further confirms that the results of evaluation are not stable, but are influenced by the dimension of the best products subset, as identified by the universities and evaluated by the panels. This is a result of the variance in efficiency of research staff within the same university: the greater this variance, the more sensitive the rankings will be to product share.

What, then, is the appropriate selection rate? The answer could be sought in optimizing the trade-off in cost versus robustness, considering that both increase with the dimension of the subset of products to be evaluated. However, the analyses conducted by the authors demonstrate that the identification of a single optimal rate would not be a straightforward task. Among additional considerations, lower selection rates not only increase the volatility in rankings, but also the probability of error in the process of selecting the best products on the part of the universities. As the literature indicates, the distribution of quality in scientific products for any given discipline is doubtless very skewed: this means that with lesser numbers of products to select it will also be more difficult to identify the differential in the best. These increasing risks of error impact negatively on the final rankings and their capacity to represent the true value of the single organizations that are evaluated.

To meet an objective of optimizing robustness in evaluation exercises, a policy maker could consider the possibility of conducting an assessment of the entire scientific portfolios of the universities. The evaluation could then, alongside a measure of excellence, offer the measure of average quality and productivity, obtained through normalization of total output against measures of input, for example the number and the academic rank of research staff employed and the resources at their disposition. But the time and costs for realizing such an exercise, using peer review methodology, would be difficult for the policy maker to accept. In light of the recent advances in bibliometric techniques, a solution that appears acceptable to the authors would be that peer review be increasingly supported (if not completely substituted, for the hard sciences) by bibliometrics. In Italy, in terms of reliability and robustness of the measures and times and costs for execution, the advantages of bibliometric techniques of evaluation (based on the ORP database), have become difficult to deny.